\documentclass{article}

\usepackage[preprint]{neurips_2020}

\usepackage[utf8]{inputenc} 
\usepackage[T1]{fontenc}    
\usepackage{hyperref}       
\usepackage{url}            
\usepackage{booktabs}       
\usepackage{amsfonts}       
\usepackage{nicefrac}       
\usepackage{microtype}      

\usepackage{algorithm}
\usepackage{algorithmic}
\usepackage{graphicx}
\usepackage{xcolor}
\usepackage{comment}
\usepackage{caption}
\usepackage{subcaption}

\usepackage{amsmath,amsfonts,bm}

\def\eqref#1{equation~\ref{#1}}

\def\1{\bm{1}}

\def\rva{{\mathbf{a}}}
\def\rvb{{\mathbf{b}}}

\def\rvf{{\mathbf{f}}}

\def\rvm{{\mathbf{m}}}

\def\rvo{{\mathbf{o}}}

\def\rvr{{\mathbf{r}}}

\def\rvx{{\mathbf{x}}}

\def\rvz{{\mathbf{z}}}

\def\rmA{{\mathbf{A}}}

\def\rmX{{\mathbf{X}}}

\DeclareMathAlphabet{\mathsfit}{\encodingdefault}{\sfdefault}{m}{sl}
\SetMathAlphabet{\mathsfit}{bold}{\encodingdefault}{\sfdefault}{bx}{n}

\newcommand{\E}{\mathbb{E}}

\newcommand{\KL}{D_{\mathrm{KL}}}

\DeclareMathOperator*{\argmax}{arg\,max}

\usepackage[greek, english]{babel}
\usepackage{textgreek}
\babeltags{gr = greek}

\usepackage{caption}
\usepackage{subcaption}

\usepackage{tikz}
\usetikzlibrary{bayesnet}

\title{Disentangling Domain and Content}

\author{%
  Dan Andrei Iliescu\thanks{} \\
  University of Cambridge\\
  \texttt{dai24@cam.ac.uk} \\
  \And
  Aliaksei Mikhailiuk \\
  University of Cambridge\\
  \texttt{am2442@cam.ac.uk} \\
  \And
  Damon Wischik \\
  University of Cambridge\\
  \texttt{djw1005@cam.ac.uk} \\
  \And
  Rafal Mantiuk \\
  University of Cambridge\\
  \texttt{rkm38@cam.ac.uk} \\
}

\begin{document}

\maketitle

\begin{abstract}
Many real-world datasets can be divided into groups according to certain salient features (e.g. grouping images by subject, grouping text by font, etc.). Often, machine learning tasks require that these features be represented separately from those manifesting independently of the grouping. For example, image translation entails changing the style of an image while preserving its content. We formalize these two kinds of attributes as two complementary generative factors called ``domain'' and ``content'', and address the problem of disentangling them in a fully unsupervised way. To achieve this, we propose a principled, generalizable probabilistic model inspired by the Variational Autoencoder. Our model exhibits state-of-the-art performance on the composite task of generating images by combining the domain of one input with the content of another. Distinctively, it can perform this task in a few-shot, unsupervised manner, without being provided with explicit labelling for either domain or content. The disentangled representations are learned through the combination of a group-wise encoder and a novel domain-confusion loss.
\end{abstract}

\section{Introduction}

Learning rich, interpretable representations with deep neural networks is one of the main challenges of current artificial intelligence research. Achieving such representations would enable us to perform complex and highly useful operations on high-dimensional data \citep{bengio2013representation}. Perhaps the first milestone that has yet to be reached in this research is learning representations which easily factorize along the lines of recognizable human concepts. This property is called ``the disentanglement of generative factors'', and is an accelerating field of inquiry \citep{tschannen2018recent}, with many major contributions coming from models based on the Variational Autoencoder \citep{kingma2013auto, rezende2014stochastic}.

Recent work \citep{locatello2018challenging,van2019disentangled} has revealed limitations in the current methods caused by the inherent ambiguity of the disentanglement objective. They have pointed out the need for equipping models with inductive biases appropriate to their respective application. An example disentanglement task with such increased specificity is the learning to generate multi-object scenes whereby the representation is trained to factorize along object lines \citep{engelcke2019genesis,burgess2019monet,greff2019multi}.

As a further step towards this goal, we identify another promising disentanglement objective, namely the separation of domain and content representations, widely applicable to a variety of tasks, ranging from unsupervised translation to missing data imputation. In a general sense, whenever there exists some form of grouping imposed on a dataset, the notion of domain arises naturally to characterize the attributes of the data which are common within groups but differ across groups. Such attributes could be the style of a painting in the context of style transfer, or the language of a text in the context of neural machine translation. The notion of content then appears as a counterpart to the domain to encompass the features which occur independently of the domain features. For instance, in the context of style transfer, the actual subject of the painting represents the content.

\subsection{Related Work}

There are many research directions which lead into the domain-content paradigm. Early work on domain adaptation \citep{ben2010theory,ganin2016domain}, for example, has highlighted the desirability of learning domain-invariant (content) representations of the data, in order to perform classification and regression in a common space. The model of \citet{gonzalez2018image} can successfully separate domain-specific from domain-invariant features for two domains.

Problems such as image-to-image translation, which entails changing the domain of an image while preserving its content, have been extensively studied. Major deep learning innovations have come from this area \citep{isola2017image, zhu2017unpaired}, producing results of excellent quality. However, unsupervised models have been limited by the rigidity of their domain representations. Many methods can only be trained to map between two domains \citep{zhu2017unpaired, taigman2016unsupervised}, or a fixed set of domains \citep{choi2018stargan, choi2019starganv2, lee2019drit++}. Even models designed to accommodate new domains at test-time either rely on restricting the domain to stylistic features \citep{liu2019few}, or requiring re-training for every new example \citep{benaim2018one}. Moreover, to the best of our knowledge, no model has the capacity to process sets of examples specifying both the source and target domain at test-time, but rely either on explicit conditioning or on one single example. All these constraints limit the model's ability to understand unseen domains and transfer knowledge between them.

Conversely, state-of-the-art methods to perform novel view synthesis rely either on structural assumptions about the geometry of the scene \citep{sitzmann2019scene, yoon2020novel} or on explicit conditioning on camera viewpoint (content) \citep{eslami2018neural, mildenhall2020nerf}. This restricts the model's usefulness when the viewpoint or scene structure is missing or difficult to describe explicitly.

Perhaps the closest inspiration for our approach comes from the influential work on Semi-Supervised learning by \citet{kingma2014semi}, where they design a generative model with two latent variables: class (domain) and $\rvz$ (content). They recognize that a limitation of their method is that the number of generative likelihood evaluations scales linearly with the number of classes, since the class variable is categorical. 

\subsection{Our contribution}

We wish to address a general formulation of the domain-content problem that does not rely on any explicit conditioning or constraints on the nature of the disentangled features. We treat domain and content as independent continuous random variables. The strength of our model is that it allows for the specification of new domains at test-time by provision of any number of examples, and permits content queries through examples as well. The continuous nature of the latent representations is also useful for measuring similarity between domains, or for classifying inputs by content irrespective of domain.

\textbf{In this work:}

\begin{itemize}
\item We propose a probabilistic model of domain-content with an associated neural architecture built upon the paradigm of the Variational Autoencoder. Our model has the capacity to separate domain and content features in an unsupervised, few-shot manner. The group-wise domain encoder enables it to process unseen domains at test-time, while its novel domain-confusion loss prevents domain and content information from mixing in the latent representations during training.

\item We demonstrate qualitatively the ability of our model to perform a generalized task called domain-content fusion, bringing together image-to-image translation and novel view synthesis, that requires the model to generate images by combining the domain of one input with the content of another.

\item We measure quantitatively the robustness of our disentanglement by ascertaining how well our model's latent representations can predict domain and content features in the data. We record improvements over other disentanglement methods.
\end{itemize}

\section{Probabilistic Domain-Content}

\begin{figure}
\centering
\includegraphics[width=\textwidth]{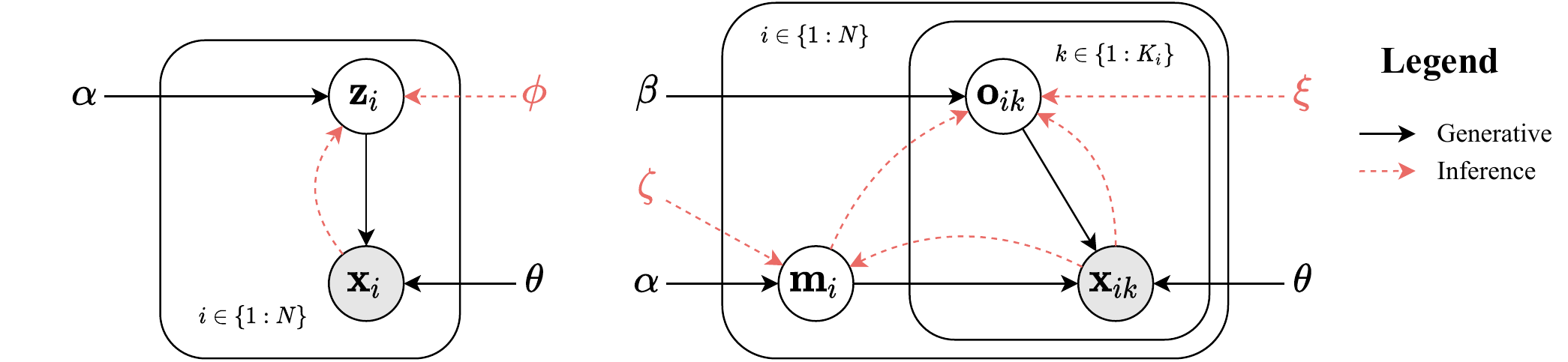}
\caption{Bayesian networks comparing VAE (left) and our Domain-Content model (right).}
\label{fig:bayes_net}
\end{figure}

We consider a very general formulation of the domain-content problem: Let $\rmX$ be a dataset of images divided into $N$ packs $\underline{\rvx}_i$ (we use the underline notation to denote a pack of elements), where $i \in \{1:N\}$. Each pack $i$ consists of $K_i$ images $\underline{\rvx}_i = \{\rvx_{ik} ~|~ k \in \{1:K_i\}\}$. By definition, all elements within a pack belong to the same domain.

Our goal is to create a probabilistic model that exploits the pack structure of the dataset in order to extract a domain random variable $\rvm_i$ and a pack of content random variables $\underline{\rvo}_i$ from a given pack of images $\underline{\rvx}_i$. In order to constitute useful disentangled representations, the inferred latent variables should satisfy certain intuitive principles:

\begin{enumerate}
\item The representation should be rich enough that one could recover an accurate estimate of the input $\underline{\rvx}_i$ given the distribution of the latent variables $\rvm_i$ and $\underline{\rvo}_i$. \label{princ:rec}
\item Each image in the pack $\underline{\rvx}_i$ should have the same associated domain variable $\rvm_i$ (we have enforced this by construction). \label{princ:dom}
\item The distribution of individual content variables $\rvo_{ik}$ should be independent of the domain of the input. \label{princ:con}
\end{enumerate}

In this work, we propose a probabilistic model with an associated neural architecture that, by following these principles, achieves effective and robust disentanglement of the domain and content factors. Our model follows the Variational Autoencoder paradigm \citep{kingma2013auto,rezende2014stochastic}, wherein the latent inference density is used as a sampling distribution for training a generative model. We have been particularly inspired by the semi-supervised approach of \citet{kingma2014semi}, who have also built a Variational Autoencoder with two latent variables.

\subsection{Parametric Generative Model}

Our generative model comprises a family of parametric densities $p$ over the variables $\underline{\rvx}_i$, $\rvm_i$ and $\underline{\rvo}_i$. The joint distribution of a pack factorizes according to the Bayesian network in Figure \ref{fig:bayes_net}:

\begin{equation}
p(\underline{\rvx}_i, \rvm_i, \underline{\rvo}_i) = p(\rvm_i) \prod_{k=1}^{K_i} p(\rvo_{ik}) p(\rvx_{ik} | \rvm_i, \rvo_{ik})
\end{equation}

Notice that, when conditioned on $\rvm_i$, an individual image variable $\rvx_{ik}$ is independent of all the other images in the pack $\underline{\rvx}_i \setminus \{ \rvx_{ik} \}$ and their corresponding contents $\underline{\rvo}_i \setminus \{ \rvo_{ik} \}$. We assign parameters $\alpha$ to the prior over the domain $p_\alpha (\rvm_i)$, $\beta$ to the prior over the content $p_\beta (\rvo_{ik})$ and $\theta$ to the generator density $p_\theta (\rvx_{ik} | \rvm_i, \rvo_{ik})$. Only $\theta$ is a trainable parameter, since it corresponds to the parameters of our decoder network. Its maximum likelihood estimator is:

\begin{align}
\theta' &= \underset{\theta}{\argmax} \frac{1}{N} \sum_{i=1}^N \log p_{\theta, \alpha, \beta} (\underline{\rvx}_i) \textrm{, where } \\ p_{\theta, \alpha, \beta} (\underline{\rvx}_i) &= \E_{p_\alpha (\rvm_i)} \prod_{k=1}^{K_i} \E_{p_\beta (\rvo_{ik})} \log p_\theta (\rvx_{ik} | \rvm_i, \rvo_{ik})    
\end{align}

\subsection{Variational Inference in the Domain-Content Model}

\begin{figure}
\centering
\hspace{-1.2cm}
\includegraphics[height=4.8cm]{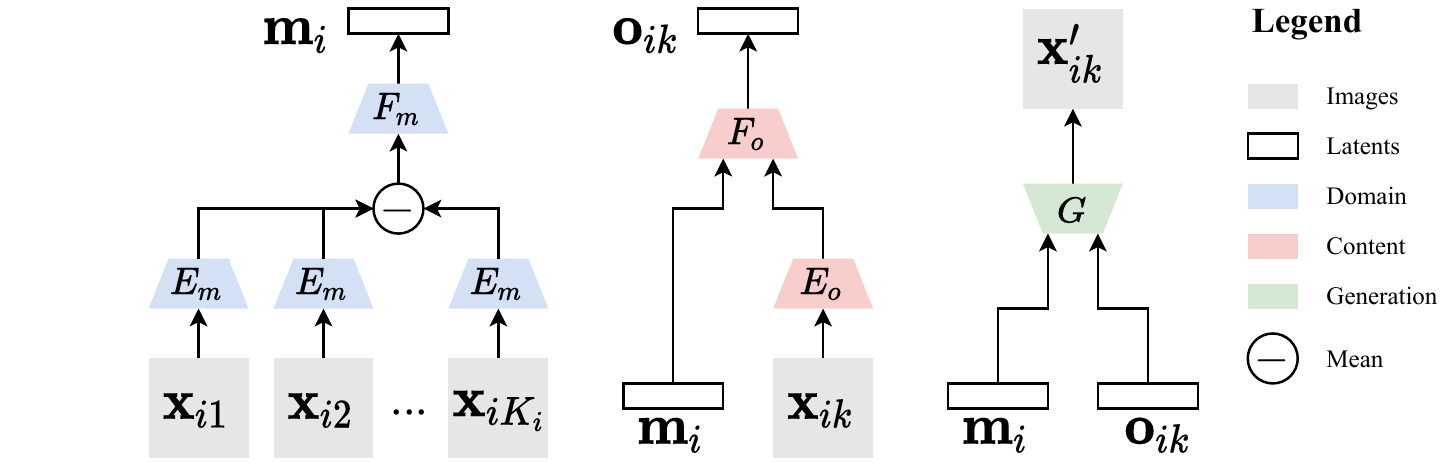}
\caption{Neural architecture of the MO model.}
\label{fig:architecture}
\end{figure}

Optimizing the likelihood under this formulation would require sampling over the priors $p_\alpha (\rvm_i)$ and $p_\beta (\rvo_{ik})$, a procedure that would converge extremely slowly and leave us with no tractable posterior over the latents conditioned on the images. We, therefore, introduce a parametric inference density $q(\rvm_i, \underline{\rvo}_i | \underline{\rvx}_i)$ over which to perform importance sampling during training for faster convergence and a tractable inference posterior. This inference density comprises the domain-content extractor that we set out to create.

According to \citet{kahn1953methods}, the generative latent posterior $p_{\theta, \alpha, \beta} (\rvm_i, \underline{\rvo}_i | \underline{\rvx}_i)$ is, itself, the optimal inference density with respect to reducing the variance of the maximum likelihood estimator. We seek, therefore, to design an inference density $q(\rvm_i, \underline{\rvo}_i | \underline{\rvx}_i)$ that preserves the conditional relationships between the variables in the generative model, in order to be theoretically capable of recovering the generative posterior. A range of choices are available on how to factorize the inference posterior while still retaining the aforementioned conditional relationships. We choose to condition the content on the domain in order to exploit the conditional independence of contents in a pack given its domain $q(\underline{\rvo}_i | \rvm_i, \underline{\rvx}_i) = \prod_{k=1}^{K_i} q(\rvo_{ik} | \rvm_i, \rvx_{ik})$. The inference model, depicted in Figure \ref{fig:bayes_net}, becomes:

\begin{equation}
q(\rvm_i, \underline{\rvo}_i | \underline{\rvx}_i) = q(\rvm_i | \underline{\rvx}_i) \prod_{k=1}^{K_i} q(\rvo_{ik} | \rvx_{ik}, \rvm_i)
\end{equation}

We assign parameters $\zeta$ to the inference posterior of the domain $q_\zeta (\rvm_i | \underline{\rvx}_i)$, and $\xi$ to the posterior inference over the content $q_\xi (\rvo_{ik} | \rvx_{ik}, \rvm_i)$. Both $\zeta$ and $\xi$ are trainable parameters, since the goal is to learn an inference density accurate enough to recover the image input (Principle \ref{princ:rec}). They correspond to the parameters of our domain and content encoder.

By applying Importance Sampling, followed by Jensen's Inequality, to the maximum likelihood objective, we obtain the Evidence Lower Bound (ELBO) for our model (the full derivation is available in Appendix A):

\begin{align}
\label{eq:elbo}
\log p_{\theta, \alpha, \beta} (\underline{\rvx}_i) \geq ~ & \E_{q_\zeta (\rvm_i | \underline{\rvx}_i)} \sum_{k=1}^{K_i} \E_{q_\xi (\rvo_{ik} | \rvx_{ik}, \rvm_i)} \log p_\theta (\rvx_{ik} | \rvm_i, \rvo_{ik}) ~~~~~~ \textrm{(reconstruction)} \\ & - \KL [q_\zeta (\rvm_i | \underline{\rvx}_i) ~||~ p_\alpha (\rvm_i)] ~~~~~~~~~~~~~~~~~~~~~~~~~~~~~~~~~~~~~~ \textrm{(domain)} \\ & - \E_{q_\zeta (\rvm_i | \underline{\rvx}_i)} \sum_{k=1}^{K_i} \KL [q_\xi (\rvo_{ik} | \rvx_{ik}, \rvm_i) ~||~ p_\beta (\rvo_{ik})] ~~~~~ \textrm{(content)}
\end{align}

This separates neatly into a reconstruction loss and two regularization penalties, for the domain and content variables. The expression of the reconstruction loss optimizes Principle \ref{princ:rec} directly, as it encourages precise estimates of the output image.

\subsection{Neural Architecture}

\begin{figure}
\begin{minipage}{0.5\textwidth}
\begin{algorithm}[H]
\caption{MorphOus}
\label{alg:model}
\begin{algorithmic}
\STATE {\bfseries Input:} Dataset $\rmX$ formed of $N$ packs each of $K_i$ images, hyperparameter $\lambda = 100$.
\FOR{$i=1$ {\bfseries to} $N$}
    \STATE Randomly select two packs $\underline{\rvx}_i$ and $\underline{\rvx}_j$
    \STATE $\rvm_i \sim q_\zeta (\rvm_i | \underline{\rvx}_i)$
    \STATE $L^m_i := \log q_\zeta (\rvm_i | \underline{\rvx}_i) - \log p_\alpha (\rvm_i)$
    \FOR{$k=1$ {\bfseries to} $K_i$}
        \STATE $\rvo_{ik} \sim q_\xi (\rvo_{ik} | \rvx_{ik}, \rvm_i)$
        \STATE $L^o_{ik} := \log q_\xi (\rvo_{ik} | \rvx_{ik}, \rvm_i) - \log p_\beta (\rvo_{ik})$
        \STATE $L^r_{ik} := (\rvx_{ik} - G(\rvm_i, \rvo_{ik}))^2$
    \ENDFOR
    \STATE $L_i = L^m_i + \sum_{k=1}^{K_i} [L^r_{ik} + L^o_{ik}]$
    \STATE \textsl{/ / repeat for pack $\underline{\rvx}_j$}
    \STATE $L_j= L^m_j + \sum_{k=1}^{K_j} [L^r_{jk} + L^o_{jk}]$
    \STATE \textsl{/ / domain-confusion loss}
    \STATE $L^c := \textrm{DomConfLoss}(\underline{\rvo}_i, \underline{\rvo}_j)$
    \STATE $L := L_i + L_j + \lambda L^c$
    \STATE \textsl{/ / update parameters}
    \STATE $\theta, \zeta, \xi := \textrm{Adam} (L)$
\ENDFOR
\end{algorithmic}
\end{algorithm}
\end{minipage}
\hspace{0.02 \textwidth}
\begin{minipage}{0.45\textwidth}
\begin{algorithm}[H]
\caption{DomConfLoss}
\label{alg:adv}
\begin{algorithmic}
\STATE {\bfseries Input:} Two packs of content variables $\underline{\rvo}_i, \underline{\rvo}_j$ of size $K_i$ and $K_j$.
\STATE {\bfseries Output:} Loss value verifying whether the packs have the same distribution.
\STATE \textsl{/ / choose random sizes for pack splits}
\STATE $A_i \sim \mathcal{U}(1, K_i), ~ B_i := K_i - A_i$
\STATE $A_j \sim \mathcal{U}(1, K_j), ~ B_j := K_j - A_j$
\STATE $\underline{\rva}_{i}, \underline{\rvb}_{i} := \textrm{split} (\underline{\rvo}_i, A_i, B_i)$
\STATE $\underline{\rva}_{j}, \underline{\rvb}_{j} := \textrm{split} (\underline{\rvo}_j, A_j, B_j)$
\STATE \textsl{/ / get ``real'' discriminator predictions}
\STATE $\rvr_i := D(\sum_{k=1}^{A_i} H(\rva_{ik}), \sum_{l=1}^{B_i} H(\rvb_{il}))$
\STATE $\rvr_j := D(\sum_{k=1}^{A_j} H(\rva_{jk}), \sum_{l=1}^{B_j} H(\rvb_{jl}))$
\STATE \textsl{/ / get ``fake'' discriminator predictions}
\STATE $\rvf_a := D(\sum_{k=1}^{A_i} H(\rva_{ik}), \sum_{l=1}^{A_j} H(\rva_{jl}))$
\STATE $\rvf_b := D(\sum_{k=1}^{B_i} H(\rvb_{ik}), \sum_{l=1}^{B_j} H(\rvb_{jl}))$
\STATE $L^c := \log \rvr_i \rvr_j + \log (1 - \rvf_a) (1 - \rvf_b)$
\STATE \textsl{/ / update discriminator parameters}
\STATE $D, H := \textrm{Adam} (- L^c)$
\STATE \textbf{return} $L^c$
\end{algorithmic}
\end{algorithm}
\end{minipage}
\end{figure}

Following the VAE paradigm, we implement the three trainable parametric densities of our model as three normal distributions with diagonal covariance, whose mean and variance are computed by feed-forward neural architectures. The generator density takes the form of a normal distribution with fixed variance and mean computed by the generator network $G$, taking as input the concatenated domain and content codes. In practice, the output image will be the mean of the distribution, rather than samples from it.

\begin{equation}
p_\theta (\rvx_{ik} | \rvm_i, \rvo_{ik}) = \mathcal{N}(\mu_x, 1) \textrm{, where } ~ \mu_x = G(\rvm_i, \rvo_{ik})    
\end{equation}

The domain inference density is a normal that requires its parameters to be computed by a neural architecture processing a variable number of un-ordered, exchangeable inputs. For this, we use a Deep Set network architecture \citep{zaheer2017deep}, whereby each input is individually encoded by a the same network $E_m$, then the outputs are averaged together, and the result is passed through a second network $F_m$. We average the outputs instead of summing them (as used in \citet{zaheer2017deep}) because we want the inference density of the domain to be agnostic to the number of inputs in the pack.

\begin{equation}
q_\zeta (\rvm_i | \underline{\rvx}_i) = \mathcal{N}(\mu_m, \sigma_m^2) \textrm{, where } ~ \mu_m, \sigma_m = F_m \left( \frac{1}{K_i} \sum_{k=1}^{K_i} E_m(\rvx_{ik}) \right)  
\end{equation}

The content inference density is a normal with parameters computed by encoding an image with a network $E_o$, then concatenating the output with the domain code of the pack, then passing it through another network $F_o$.

\begin{equation}
q_\xi (\rvo_{ik} | \rvx_{ik}, \rvm_i) = \mathcal{N}(\mu_o, \sigma_o^2) \textrm{, where } ~ \mu_o, \sigma_o = F_o(\rvm_i, E_o(\rvx_{ik}))  
\end{equation}

Diagrams depicting each of these architectures are displayed in Figure \ref{fig:architecture}. We employ the reparametrization trick \citep{kingma2013auto} to sample from the inference posterior over the latents. As for the domain and content priors, they are fixed arbitrary normal distributions with mean 0 and variance 1. A more complete specifications of the neural implementation is available in Appendix B.

\section{Domain-Confusion Loss}

\begin{figure}
\centering
\includegraphics[height=4.8cm]{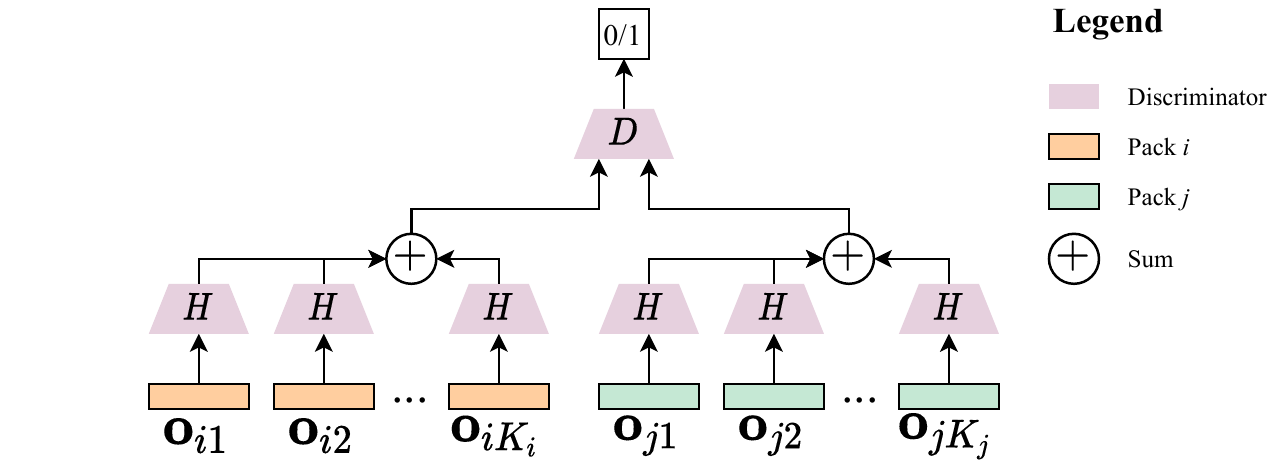}
\caption{Neural architecture of the Domain-Confusion discriminator $\eta$.}
\label{fig:discriminator}
\end{figure}

So far, Principle \ref{princ:rec} is encouraged by optimizing the Evidence Lower Bound, while Princicple \ref{princ:dom} is enforced by construction. We inspect whether the model also satisfies Principle \ref{princ:con}, which requires that the distribution of inferred content variables in a pack be orthogonal to the domain of the pack. We can reformulate this principle more precisely to claim that, in the limit of infinitely large packs, the distribution of a random variable $\tilde{\rvo}$, denoting the random choice of one content variable $\rvo_{ik}$ from a pack of inferred content variables $\underline{\rvo}_i$, should be the same same regardless of the pack of origin.

\begin{equation}
\label{eq:choose}
q_{\zeta, \xi}(\tilde{\rvo} | \underline{\rvx}_i) = q_{\zeta, \xi}(\tilde{\rvo} | \underline{\rvx}_j) \textrm{, where } K_i, K_j \to \infty
\end{equation}

This statement relies on the fact that, when the size of the pack tends to infinity, the random picking of one inferred content variable from the pack defines a distribution over content values conditioned on the ``true'' domain of the pack. Since the probability of content features should be independent of domain features, the empirical density $q(\tilde{\rvo} | \underline{\rvx}_i)$ should also stay fixed as $i$ changes.

\paragraph{Claim} When the Domain-Content Evidence Lower Bound (equation \ref{eq:elbo}) is maximized to its theoretical potential, then equation \ref{eq:choose} is satisfied. In other words, when the inference latent posterior $q_{\zeta, \xi}(\rvm_i, \underline{\rvo}_i | \underline{\rvx}_i)$ approaches the generative latent posterior $p_{\theta, \alpha, \beta}(\rvm_i, \underline{\rvo}_i | \underline{\rvx}_i)$, and the generative data likelihood $p_{\theta, \alpha, \beta} (\underline{\rvx}_i)$ produces samples indistinguishable from the real data, then the distribution of content variables will become independent of the ``true'' domain of the pack, in the limit of large packs. A discussion and proof of this is included in Appendix A.

This result shows that our probabilistic model is, in theory, sufficient to satisfy Principle \ref{princ:con}. However, this state cannot be achieved in practice, because of architectural limitations on both the generative and inference density families. Therefore, in order to encourage the realization of Principle \ref{princ:con}, we can constrain the space of inference densities to those for which equation \ref{eq:choose} holds, at least approximately. This need is reinforced by empirical observations of the unconstrained model, which reveal that the distribution of inferred content variables within a pack is highly sensitive to the task, network architecture and choice of hyperparameters.

In this work, we tackle this problem practically by proposing an adversarial loss which encourages the homogeneous distribution of content variables across packs by penalizing differences between pairs of packs in their set of content values. We call this the Domain-Confusion loss, and we show empirically in Table \ref{tab:pred} that it increases the quality and robustness of the disentanglement. 

The loss is built around an adversarial discriminator $\eta$ which receives as input a pair of packs of content values and is trained to output 1 if the two packs come from the same distribution, or 0 if they are differently distributed. This is called a verification task, inspired by \citet{sohn2018unsupervised}, and we apply it to contrast pairs of sub-packs coming from the same pack with pairs of sub-packs coming from different packs. Concretely, every iteration of training takes as input two packs of content values $\underline{\rvo}_i, \underline{\rvo}_j$, which we split randomly into $(\underline{\rva}_i, \underline{\rvb}_i)$ and $(\underline{\rva}_j,\underline{\rvb}_j)$, respectively. The loss takes the value:

\begin{equation}
L^c (\underline{\rvo}_i, \underline{\rvo}_j) = \log \eta(\underline{\rva}_i, \underline{\rvb}_i) + \log \eta(\underline{\rva}_j, \underline{\rvb}_j) + \log (1 -  \eta(\underline{\rva}_i, \underline{\rva}_j)) + \log (1 -  \eta(\underline{\rvb}_i, \underline{\rvb}_j))
\end{equation}

The greater the loss, the more the discriminator can distinguish between the two packs. Because the architecture of $\eta$ needs to accommodate two packs of varying sizes, we implement it as a Deep Set \citep{zaheer2017deep}, just like in the case of the domain encoder. Unlike to the domain encoder, here we use summation instead of averaging, since $\eta$ does not need to be agnostic to the number of inputs. The discriminator takes the form: 

\begin{equation}
\eta (\underline{\rvo}_i, \underline{\rvo}_j) = D \left( \sum_{k=1}^{K_i} H(\rvo_{ik}), ~ \sum_{l=1}^{K_j} H(\rvo_{jl}) \right)
\end{equation}

where $D$ and $H$ are neural networks. A diagram of this architecture is displayed in Figure \ref{fig:discriminator}. The full implementation is available in Algorithm \ref{alg:adv}.

\section{Experiments}
\label{sec:exp}

\begin{figure}
\begin{subfigure}[b]{0.49\textwidth}
\centering
\includegraphics[width=\textwidth]{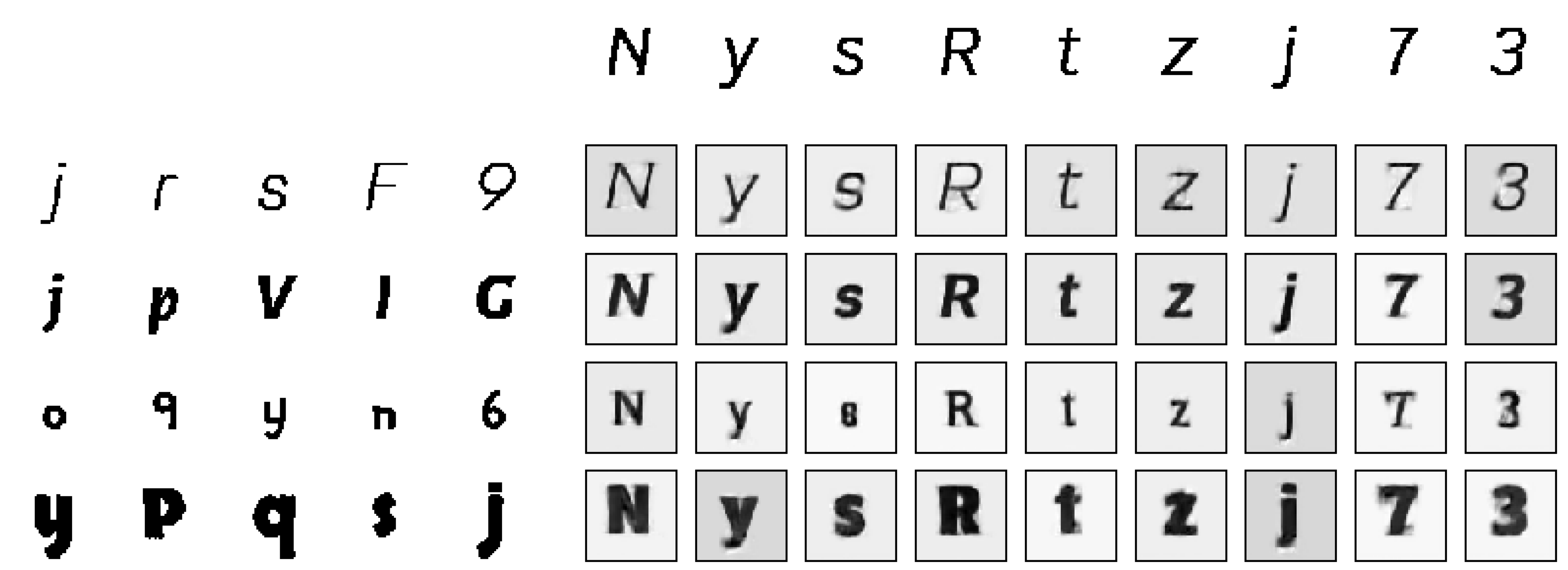}
\caption{Disentangling font and character.}
\label{fig:font}
\end{subfigure}
\hspace{0.02\textwidth}
\begin{subfigure}[b]{0.49\textwidth}
\centering
\includegraphics[width=\textwidth]{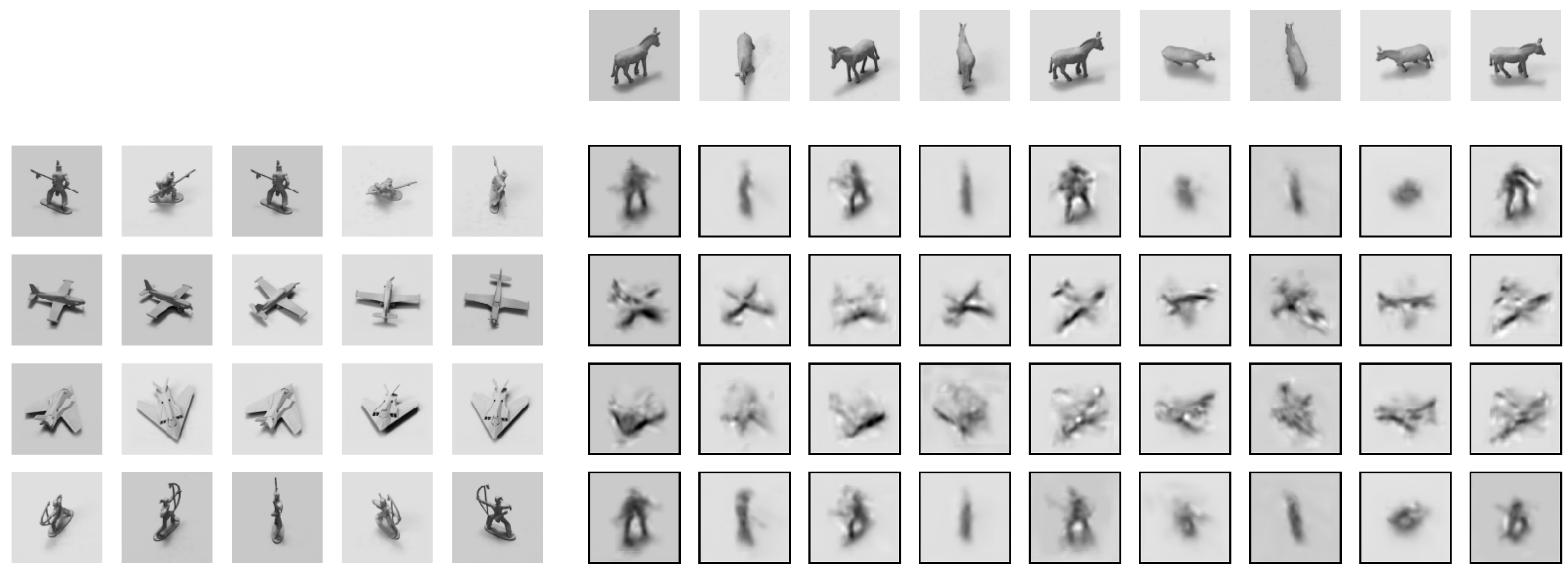}
\caption{Disentangling object and imaging conditions.}
\label{fig:norb}
\end{subfigure}
\\\\
\begin{subfigure}[b]{1.\textwidth}
\centering
\includegraphics[width=\textwidth]{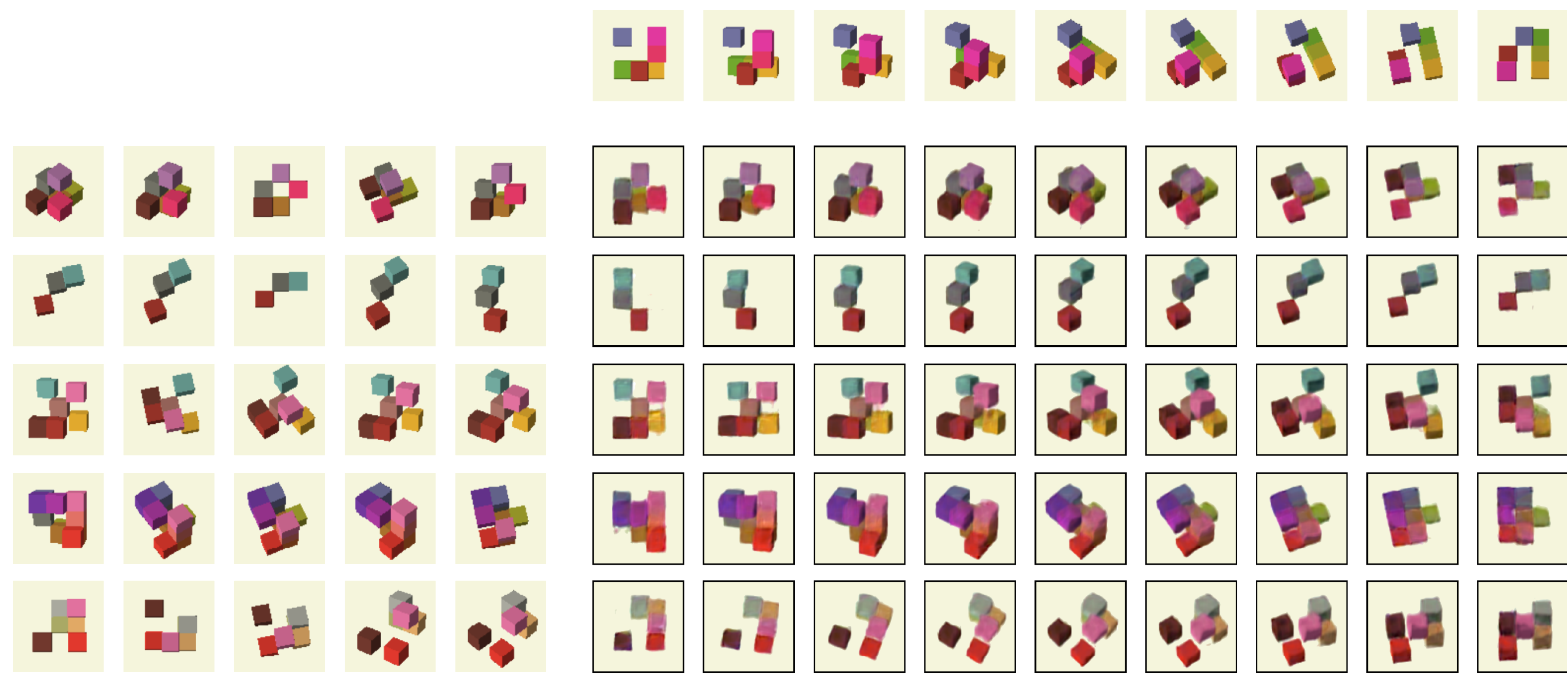}
\caption{Disentangling object shape and object rotation.}
\label{fig:silhouettes}
\end{subfigure}
\\\\
\caption{Qualitative results of our model on the fusion task are displayed with a black border. Each row represents a domain and each column represents a content. The topmost row is the content input and each row of the leftmost 5 columns is a domain input.}
\label{fig:res}
\end{figure}

We evaluate our model on the generalized domain-content fusion task mentioned in the introduction. Given a trained model and two unseen image packs $\underline{\rvx}_i$, $\underline{\rvx}_j$, we extract the domain of $\underline{\rvx}_i$ and the contents of every image in $\underline{\rvx}_j$, and then use them to generate $K_j$ output images.

This task enables us to visually inspect the quality of the disentanglement by judging how well the model follows each of the domain-content principles: Are the images of high-quality? Do they have the same domain features? Do the content features of each correspond to those of the associated input image?

We apply our model to three datasets: a dataset of font images collected from Google Fonts, the Small Norb Dataset \citep{lecun2004learning} and an original dataset, called Silhouettes, comprising 3-dimensional block shapes imaged at various rotation angles. Detailed descriptions of the datasets are available in Appendix C. For the Silhouettes and Google Font datasets, we test only on domains which have been withheld during training, in order to show how well the model generalizes to new domains. We display results for 5 testing packs of each dataset in Figure \ref{fig:res}. Further results on more datasets are available in the Appendix D. We also provide results in Appendix D for testing the model on both unseen domains and unseen contents. The results appear to separate domain and content features very well without sacrificing the image quality as compared to the VAE.

\subsection{Predicting ground-truth factors from the latent representation}

In order to obtain more quantitative evidence of disentanglement, we adapt the factor regression method also used by \citet{greff2019multi} to measure factor information and interpretability in the latent space. The method involves learning a simple linear mapping between the latent space of the trained model and the value space of the ground-truth factor. A high predictive accuracy implies that the latent representation contains the necessary factor information and also organizes it in an easily interpretable way.

We complete this experiment on our Silhouettes dataset, and learn separate predictors for the domain features (object shape) and content features (rotation angle). Predicting the shape is a 27-way binary classification and predicting the rotation angle is a two-way regression. We provide more details on the encoding of these features in Appendix C. We compare each latent representation of our model (domain and content) with a VAE, a FactorVAE \citep{kim2018disentangling}, considered to be state-of-the-art in disentanglement, and random guessing. In the case of our model, the goal is that each of the two representations should predict its own factor as much as possible, and not to predict the other's factor. We include a comparison of our model with and without the Domain-Confusion loss as an ablation study on the impact of this loss. Details on the experimental setup and measurements, as well as comparisons with other models on other datasets, are available in Appendix D.

\begin{table}
  \caption{How well can the latent representation predict the ground-truth factors}
  \label{tab:pred}
  \centering
  \begin{tabular}{llllllll}
    \toprule
    & \multicolumn{2}{c}{MO (w/ DC)} & \multicolumn{2}{c}{MO (w/o DC)} \\
    \cmidrule(r){2-5}
    Factor (metric) & domain & content & domain & content & FactorVAE & VAE & Guessing \\
    \midrule
    Object shape (CE) & \textbf{-0.023} & -0.449 & -0.051 & -0.402 & -0.211 & -0.236 & -0.451 \\
    Rotation (MSE) & 673.4 & \textbf{456.2} & 672.8 & 533.2 & 563.1 & 597.8 & 671.3 \\
    \bottomrule
  \end{tabular}
\end{table}

The results, displayed in Table \ref{tab:pred}, reveal not only that the latent variables of our model predict their corresponding factors far better than the FactorVAE or VAE, but also that the cross-over predictions are no better than random. This result shows that domain and content features are concentrated successfully in their designated representations. Moreover, we can see a marked improvement in the model with the Domain-Confusion loss over the one without it, in the case of both domain and content factors.

\section{Conclusion}

In this work, we have described a general problem of disentangling the domain and content generative factors, and proposed a probabilistic model with an associated neural network to solve this task. We have built the model according to the VAE paradigm and introduced the Domain-Confusion loss to compensate for limitations brought upon by the neural architecture. We have proven the effectiveness of our disentanglement solution by providing both qualitative results on the domain-content fusion tasks, and quantitative measures of the predictive power of the latent representations.

One crucial direction to explore in future work is the relationship between domain-content disentanglement and the notion of invariant risk \citet{arjovsky2019invariant} in the context of causal inference. Designing methods that can separate confounding environmental factors (domain) from the factors of interest (content) would lead to significant innovations in many research fields, especially in the study of medical counterfactuals. The method presented here is not yet able to perform such separation in the case where content distributions vary across domains, since it uses the homogeneity across domains as a proxy for identifying content factors.

\section{Broader Impact}

One of the main ethical faults associated with the application of statistical learning to real-world problems is the acquisition of any biases that might be present in the training dataset. As \citet{arjovsky2019invariant} discuss in their work, classical deep learning methods minimizing the expected risk of their hypothesis cannot distinguish between spurious correlations and true mechanisms. Therefore, naive statistical correspondences are drawn between phenomena that are not causally connected, creating unintended consequences with potential scientific or social impact. Although our model is still an expected-risk minimizing algorithm, we note that it could also be used as a paradigm for diagnosing dataset biases. For example, if two distinct populations of elements, sharing the same set of classes in a classification task, vary in their representation of different classes across different datasets, then the same individual placed in various datasets will produce, in turn, different content encodings, revealing the biases in the dataset. Our model is still very theoretical, but we believe it is a step towards a deeper study of the relationship between element and environment in the context of deep learning.

\bibliography{morphous_neurips_2020}
\bibliographystyle{icml2020}

\newpage
\section*{Appendix A - Proofs}

\subsection*{Evidence Lower Bound}

\begin{align}
\log p_{\theta, \alpha, \beta}(\underline{\rvx}_i) &= \log \E_{p_{\alpha, \beta}(\rvm_i, \underline{\rvo}_i)} p_\theta(\underline{\rvx}_i | \rvm_i, \underline{\rvo}_i) ~~~~~~ \textrm{(total probability)} \\
&= \log \E_{q_{\zeta, \xi}(\rvm_i, \underline{\rvo}_i | \underline{\rvx}_i)} \left[ p_\theta(\underline{\rvx}_i | \rvm_i, \underline{\rvo}_i) \frac{p_{\alpha, \beta}(\rvm_i, \underline{\rvo}_i)}{q_{\zeta, \xi}(\rvm_i, \underline{\rvo}_i | \underline{\rvx}_i)} \right] ~~~~~~ \textrm{(importance sampling)} \\
&\geq \underbrace{\E_{q_{\zeta, \xi}(\rvm_i, \underline{\rvo}_i | \underline{\rvx}_i)} \log p_\theta(\underline{\rvx}_i | \rvm_i, \underline{\rvo}_i)}_{\textrm{reconstruction}} + \underbrace{\E_{q_{\zeta, \xi}(\rvm_i, \underline{\rvo}_i | \underline{\rvx}_i)} \log \frac{p_{\alpha, \beta}(\rvm_i, \underline{\rvo}_i)}{q_{\zeta, \xi}(\rvm_i, \underline{\rvo}_i | \underline{\rvx}_i)}}_{\textrm{latent penalties} ~ (\rmA)} ~~~~~~ \textrm{(Jensen's inequality)}
\end{align}

Latent penalties:

\begin{align}
-\rmA =& ~ \E_{q_{\zeta, \xi}(\rvm_i, \underline{\rvo}_i | \underline{\rvx}_i)} \log \frac{q_{\zeta} (\rvm_i | \underline{\rvx}_i) \prod_{k=1}^{K_i} q_{\xi}(\rvo_{ik} | \rvx_{ik}, \rvm_i)}{p_{\alpha} (\rvm_i) \prod_{k=1}^{K_i} p_{\beta}(\rvo_{ik})} ~~~~~~ \textrm{(factorization)} \\
=& ~ \underbrace{\KL [q_{\zeta} (\rvm_i | \underline{\rvx}_i) || p_{\alpha} (\rvm_i)]}_{\textrm{domain}} + \underbrace{\E_{q_\zeta (\rvm_i | \underline{\rvx}_i)} \sum_{k=1}^{K_i} \KL [q_\xi (\rvo_{ik} | \rvx_{ik}, \rvm_i) ~||~ p_\beta (\rvo_{ik})]}_{\textrm{content}}
\end{align}

We can also rewrite the ELBO in terms of the marginal data likelihood by rearranging the terms:

\begin{equation}
ELBO(\underline{\rvx}_i) = \underbrace{\log p_{\theta, \alpha, \beta}(\underline{\rvx}_i)}_{\textrm{data likelihood}} - \underbrace{\KL[q_{\zeta, \xi}(\rvm_i, \underline{\rvo}_i | \underline{\rvx}_i) || p_{\alpha, \beta}(\rvm_i, \underline{\rvo}_i)]}_{\textrm{variational gap}}
\end{equation}

\subsection*{Content Independence}

We first introduce $\tilde{\rvx}$ which represents the image variable from input pack $\underline{\rvx}_i$ which corresponds to the selected content $\tilde{\rvo}$. $\tilde{\rvx}$ is distributed according to the empirical density $h(\tilde{\rvx} | \underline{\rvx}_i)$. We apply the law of total probability in order to obtain an expectation of the content inference density $q_\xi$ in terms of $\tilde{\rvx}$ and $\rvm_i$.

\begin{align}
q_{\zeta, \xi} (\tilde{\rvo} | \underline{\rvx}_i) &= \E_{h(\tilde{\rvx} | \underline{\rvx}_i)} q_{\zeta, \xi} (\tilde{\rvo} | \tilde{\rvx}, \underline{\rvx}_i) ~~~~~~ \textrm{(total probability)} \\
&= \E_{h(\tilde{\rvx} | \underline{\rvx}_i)} \E_{q_{\zeta}(\rvm_i | \underline{\rvx}_i)} q_{\zeta, \xi} (\tilde{\rvo} | \tilde{\rvx}, \underline{\rvx}_i, \rvm_i) ~~~~~~ \textrm{(total probability)} \\
&= \E_{h(\tilde{\rvx} | \underline{\rvx}_i)} \E_{q_{\zeta}(\rvm_i | \underline{\rvx}_i)} q_{\xi} (\tilde{\rvo} | \tilde{\rvx}, \rvm_i) ~~~~~~ \textrm{(conditional independence)}
\end{align}

As $K_i \to \infty$, selecting an image variable from the pack $h(\tilde{\rvx} | \underline{\rvx}_i)$ becomes equivalent to sampling an image variable directly from the generative density of the $p_{\theta, \beta} (\tilde{\rvx} | \rvm_i')$, where $\rvm_i'$ is the value of the ``true domain'' of pack $\underline{\rvx}_i$. For this we have also assumed that the input images come from the generative distribution $p_{\theta, \alpha, \beta} (\underline{\rvx}_i)$, one of the necessary conditions for which the ELBO is maximised.

\begin{equation}
h(\tilde{\rvx} | \underline{\rvx}_i) \overset{K_i \to \infty}{\approx} \E_{p_{\theta, \beta} (\underline{\rvx}_i | \rvm_i')} h(\tilde{\rvx} | \underline{\rvx}_i) = p_{\theta, \beta} (\tilde{\rvx} | \rvm_i')
\label{eq:img_infty}
\end{equation}

We make the same observation for the domain inference density. In order to express $\rvm_i$ in terms of $\rvm_i'$ we use the fact that the inference posterior density approaches the generative posterior density, which is the other necessary condition for which the ELBO is maximized.

\begin{equation}
q_{\zeta} (\rvm_i | \underline{\rvx}_i) \overset{K_i \to \infty}{\approx} \E_{p_{\theta, \beta} (\underline{\rvx}_i | \rvm_i')} q_{\zeta} (\rvm_i | \underline{\rvx}_i) \overset{q_{\zeta, \xi} \to p_{\theta, \alpha, \beta}}{\approx} p(\rvm_i | \rvm_i')
\label{eq:dom_infty}
\end{equation}

$p(\rvm_i | \rvm_i')$ is an artificial density which simply indicates that all the probability mass rests on the value $\rvm_i'$. We can now re-write the latent posterior as being the content posterior conditioned on the ``true domain''.

\begin{equation}
\E_{q_{\zeta}(\rvm_i | \underline{\rvx}_i)} q_{\xi} (\tilde{\rvo} | \tilde{\rvx}, \rvm_i) \overset{\textrm{Eq. \ref{eq:dom_infty}}}{\approx} \E_{p(\rvm_i, \rvm_i')} q_{\xi} (\tilde{\rvo} | \tilde{\rvx}, \rvm_i) \overset{q_{\zeta, \xi} \to p_{\theta, \alpha, \beta}}{\approx} p_{\theta, \beta} (\tilde{\rvo} | \tilde{\rvx}, \rvm_i')
\label{eq:dom_rep}
\end{equation}

We now introduce the results we have obtained into the initial expression.

\begin{equation}
q_{\zeta, \xi} (\tilde{\rvo} | \underline{\rvx}_i) \overset{Eq.\ref{eq:img_infty}, ~ \ref{eq:dom_rep}}{\approx} \E_{p_{\theta, \beta} (\tilde{\rvx} | \rvm_i')} p_{\theta, \beta} (\tilde{\rvo} | \tilde{\rvx}, \rvm_i') = p_{\beta}(\tilde{\rvo} | \rvm_i') = p_{\beta} (\tilde{\rvo}) ~~~~~~ \textrm{(cond. independence)}
\end{equation}

We have shown that, \textbf{if} $q_{\zeta, \xi} \to p_{\theta, \alpha, \beta}$, $K_i \to \infty$, and the input images are generated by $p_{\theta, \alpha, \beta}$, \textbf{then} $q_{\zeta, \xi} (\tilde{\rvo} | \underline{\rvx}_i) \approx p_{\beta} (\tilde{\rvo})$ which is independent of the pack or its domain.

\section*{Appendix B - Architectural Details}

The model consists of four networks: domain encoder, content encoder, decoder and discriminator. In this section we provide their implementation. For all networks we use Adam optimizer with learning rate of 0.0001 and $\epsilon$ of 1e-8. We set the latent code size for both domain and content to 16, we refer to them as $S_D$ and $S_C$ respectively. The encoder and decoder networks use spatial broadcasting to aid disentanglement, as introduced by \citep{watters2019spatial}.

\paragraph{Domain encoder:}
Before being fed into the encoder images are transformed into the coordinate matrix - recording the row index and column index for every pixel for every image in the pack: ($i^{th}$ pack size $K_i$)$\times$(image height $H$)$\times$(image width $W$) $\times$(image channels $C + 2 $ (one for rows, one for cols). 
The summary is provided in Table \ref{tab:architecture_dom_enc}.

The coordinate matrix is concatenated along the input pack dimension and passed through the convolutional layers. After the inputs are passed through the convolutional layers the output is reshaped into: $K_i\times(H * W/(16*16/128))$. The latent codes are then averaged along the pack dimension. The output of the operation has $1\times(H * W / 2)$ dimensions. It is then passed through a fully connected layer which maps the latent space to the mean and variance of domain distribution with shape $1\times S_D$.

\begin{table}[ht]
  \caption{Architecture of the domain encoder}
  \label{tab:architecture_dom_enc}
  \centering
  \begin{tabular}{lllll}
    \toprule
    Type & Stride & Size/Ch & Act. Fun & Comment\\
    \midrule
    Conv 4x4 & 1 & 64 & ReLU &  \\
    Conv 4x4 & 1 & 64 & ReLU &  \\
    Conv 4x4 & 4 & 128 & ReLU  &  \\
    Conv 4x4 & 4 & 128 & ReLU &  \\
    Reshape & - & - & - &   Output dims: $K_i\times(H * W/(16*16/128))$\\
    Average & - & - & - & Output dims: $1\times(H * W / 2)$ \\ 
    Dense & - & 512 & ReLU & \\ 
    Dense & - &  $K_i\times S_D$ & - & Outputs the mean of the distribution \\ 
    Dense & - &  $K_i\times S_D$ & - &  Outputs the variance of the distribution \\ 
    \bottomrule
  \end{tabular}
\end{table}

\paragraph{Content encoder:} The architecture of the content encoder follows the same general principle as domain encoder up to the averaging layer. However, now, instead of averaging along the pack we concatenate a domain code on top of every current code to get tensor of size $K_i \times (H * W/ 2 + S_D)$.

The next step is the same as with domain encoder for encoding the means and variances $K_i\times S_C$. Note that now there is only one domain code and as many content codes as there are elements in the pack. Detailed architecture is given in Table \ref{tab:architecture_con_enc}.

\begin{table}[ht]
  \caption{Architecture of the content encoder}
  \label{tab:architecture_con_enc}
  \centering
  \begin{tabular}{lllll}
    \toprule
    Type & Stride & Size/Ch & Act. Fun & Comment\\
    \midrule
    Conv 4x4 & 1 & 64 & ReLU &  \\
    Conv 4x4 & 1 & 64 & ReLU &  \\
    Conv 4x4 & 4 & 128 & ReLU  &  \\
    Conv 4x4 & 4 & 128 & ReLU &  \\
    Reshape & & & & Output dims: $K_i\times(H * W/(16*16/128))$ \\    
    Concatenate & & & & Output dims: $K_i \times (H * W/ 2 + S_D)$ \\
    Dense & - & 512 & ReLU & \\  
    Dense & - &  $K_i\times S_C$ & - & Outputs the mean of the distribution \\ 
    Dense & - &  $K_i\times S_D$ & - &  Outputs the variance of the distribution \\ 
    \bottomrule
  \end{tabular}
\end{table}

\textbf{Spatial broadcasting decoder:}
We follow the same principles as \citet{watters2019spatial}. The full architecture is given in Table \ref{tab:architecture_dec}. The decoder takes as input is domain and and content codes. These are concatenated to obtain a array of $K_i \times (S_D + S_C)$
dimensions. The array is then broadcast to the image shape 
$K_i\times H \times W \times (S_D + S_C)$. 

We then concatenate the resulting array with a coordinate matrix of size $K_i \times H \times W \times 2 $ (first channel counts the rows and second channel counts the cols). This way we end up with a array of $K_i \times H \times W \times (S_D + S_C + 2)$. Thus, for every pixel in every image in pack, we have the whole domain code, the whole content code, row number and column number.

\begin{table}[ht]
  \caption{Architecture of the decoder}
  \label{tab:architecture_dec}
  \centering
  \begin{tabular}{lllll}
    \toprule
    Type & Stride & Size/Ch & Act. Fun & Comment\\
    \midrule
    Conv 4x4 & 1 & 128 & ReLU &  \\
    Conv 4x4 & 1 & 128 & ReLU &  \\
    Conv 4x4 & 1 & 128 & ReLU &  \\
    Conv 4x4 & 1 & 64 & ReLU  &  \\
    Conv 4x4 & 1 & 64 & ReLU  &  \\
    Conv 4x4 & 1 & C & - &  Output has image shape $H \times W \times C$\\ 
    \bottomrule
  \end{tabular}
\end{table}

\textbf{Discriminator:}
The discriminator takes as input two packs A and B of content codes with sizes $K_A$ and $K_B$ respectively. The overall input size is thus $K_A \times S_C$, $K_B \times S_C$. The output of the discriminator is the probability of the content codes coming from the same domain. The detailed architecture is given in Table \ref{tab:architecture_disc}.

\begin{table}[ht]
  \caption{Architecture of the discriminator}
  \label{tab:architecture_disc}
  \centering
  \begin{tabular}{lllll}
    \toprule
    Type & Stride & Size/Ch & Act. Fun & Comment\\
    \midrule
    Dense & - & 64 & ReLU &  For each pack separately \\  
    Dense & - & 128 & ReLU &  For each pack separately\\ 
    Sum  & - & 128  & - & Sum along pack dimension \\
    Concat & - & 256 &- &  Concatenate packs along the feature axis \\
    Dense & - & 128 & ReLU &   \\ 
    Dense & - & 64 & ReLU &   \\ 
    Dense & - & 32 & ReLU &   \\ 
    Dense & - & 1 & - &  \\ 
    \bottomrule
  \end{tabular}
\end{table}

\paragraph{Domain Confusion Loss:} When adding the domain-confusion loss to the ELBO, the $\lambda$ coefficient of the domain-confusion loss is 100.

\section*{Appendix C - Datasets}
For training we generate 16000 packs. To generate a pack we first generate a set of images belonging to the same domain and then randomly sample from this set. Each pack has a variable number of examples. The size of the pack is given by $4 + Pois(8)$, where $Pois(8)$ is a Poisson distribution with $\lambda=8$ (i.e.~ on average each pack contained 12 examples).

\paragraph{Silhouettes Dataset:} The dataset consists of $96\times96\times3$ procedurally images of 3D arrangements of cubes imaged at different rotation angles. We treat the arrangement of the cubes as the domain. We generate images by first defining a $3\times 3 \times 3$ grid. Each grid cell can be empty or occupied by a cube. We set the probability of a cube being present in a cell by Bernoulli distribution with $p=\frac{1}{6}$. For each domain we treat rotation (view angle) of the grid as the content. The angle is controlled by two parameters: pitch and yaw. Both are uniformly distributed between 0 and 90. We withhold $1000$ shapes during training (of the possible $2^27$ shapes).

\paragraph{Google Fonts:} The dataset consists of $64\times64\times1$ images of 64 characters in 2767 font typefaces, downloaded from (\url{https://github.com/google/fonts}). We treat font as the domain and character as a content. During training we withhold 100 fonts chosen randomly. In the unseen-content experiments we also withhold 4 characters chosen randomly.

\paragraph{Small Norbs:} The dataset \citep{lecun2004learning} consists of $96\times96\times1$ images of toys 50 photographed under 972 conditions (camera azimuth, rotation and lighting). The toy represents the domain and the condition represents the content. We withhold 5 toys during training.

\section*{Appendix D - Experiments}

\begin{figure}
\begin{subfigure}[b]{\textwidth}
\includegraphics[height=2.55cm]{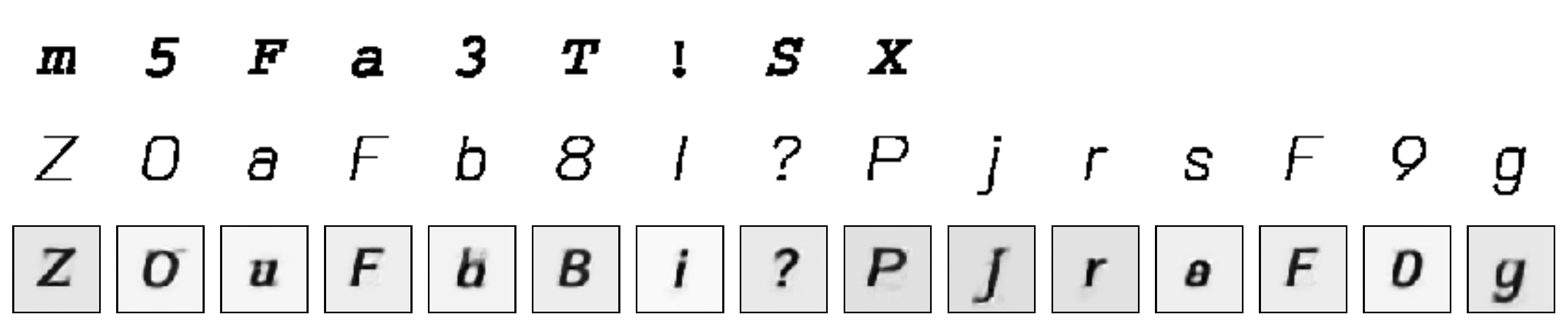}
\caption{First row is domain input, second row is content input, third row is fusion output.}
\end{subfigure}
\\
\begin{subfigure}[b]{\textwidth}
\includegraphics[height=2.55cm]{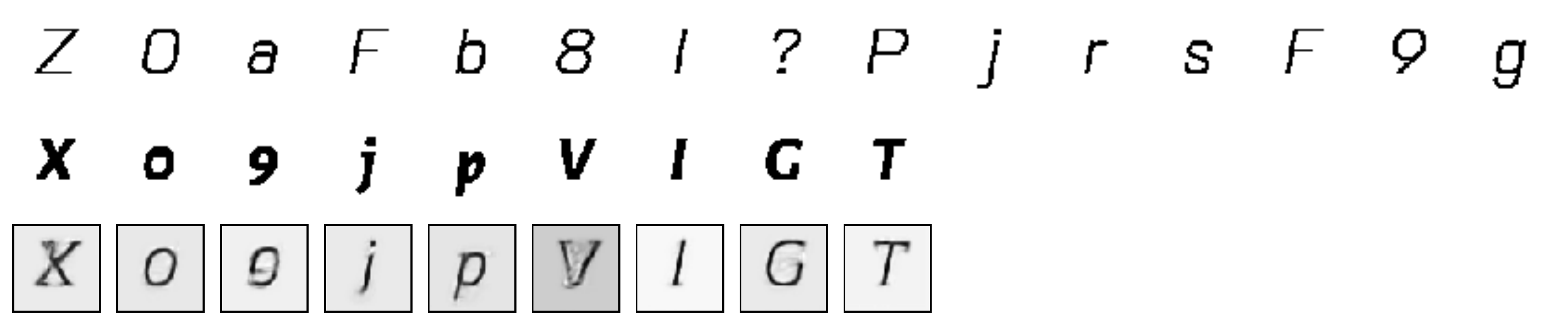}
\caption{First row is domain input, second row is content input, third row is fusion output.}
\end{subfigure}
\\
\begin{subfigure}[b]{\textwidth}
\includegraphics[height=2.55cm]{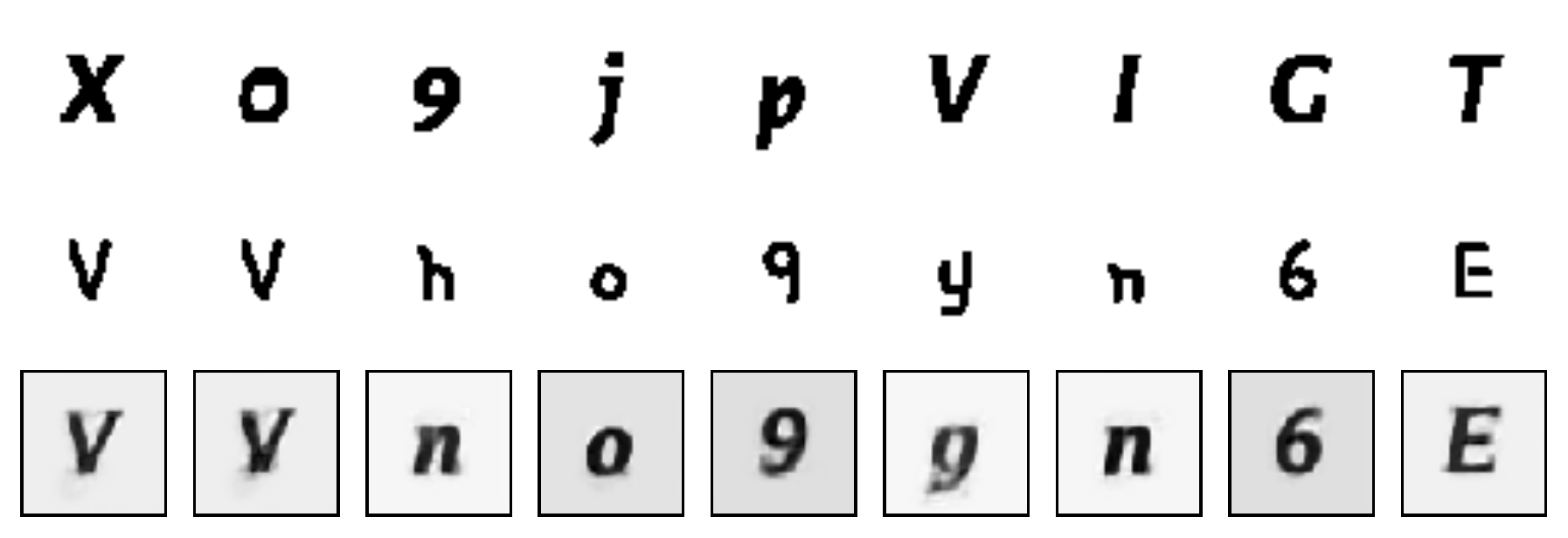}
\caption{First row is domain input, second row is content input, third row is fusion output.}
\end{subfigure}
\\
\begin{subfigure}[b]{\textwidth}
\includegraphics[height=2.55cm]{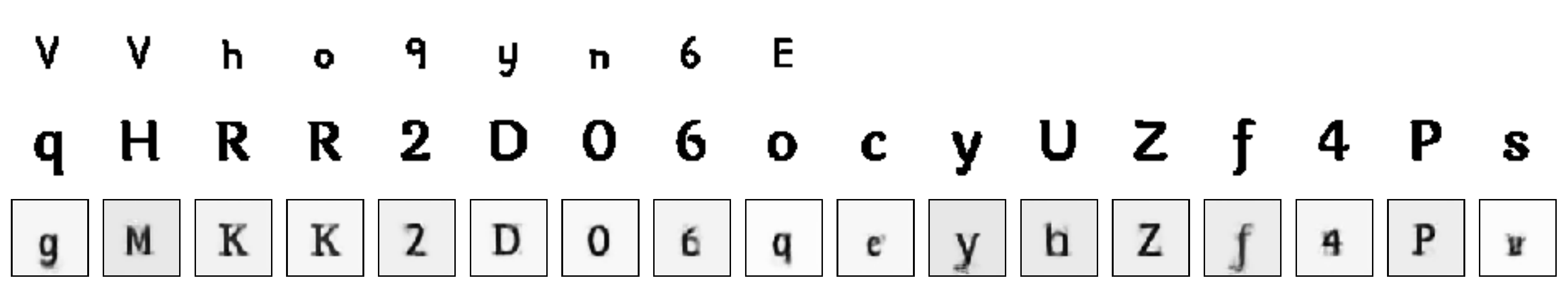}
\caption{First row is domain input, second row is content input, third row is fusion output.}
\end{subfigure}
\\
\begin{subfigure}[b]{\textwidth}
\includegraphics[height=2.55cm]{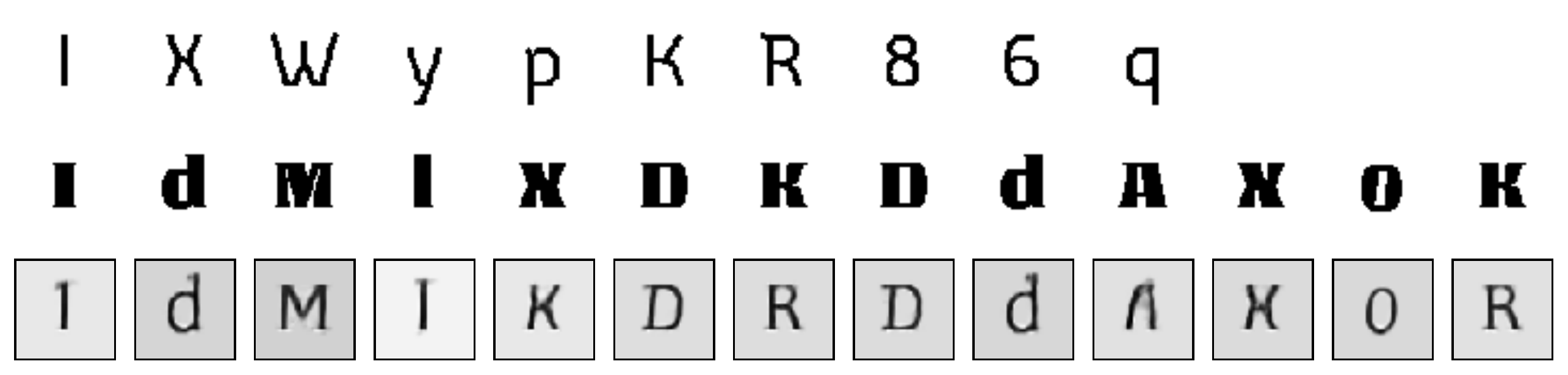}
\caption{First row is domain input, second row is content input, third row is fusion output.}
\end{subfigure}
\\
\begin{subfigure}[b]{\textwidth}
\includegraphics[height=2.55cm]{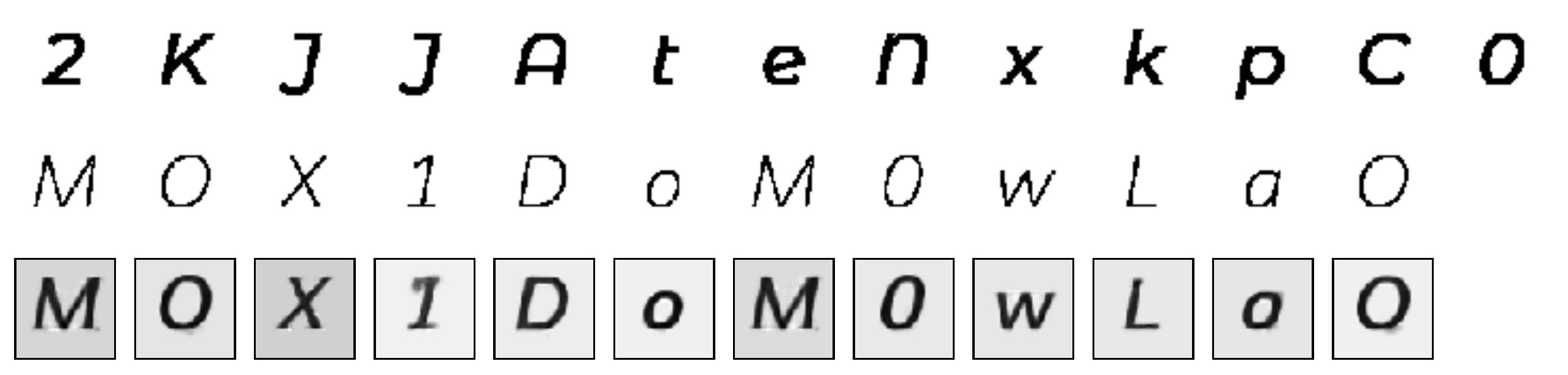}
\caption{First row is domain input, second row is content input, third row is fusion output.}
\end{subfigure}
\\
\begin{subfigure}[b]{\textwidth}
\includegraphics[height=2.55cm]{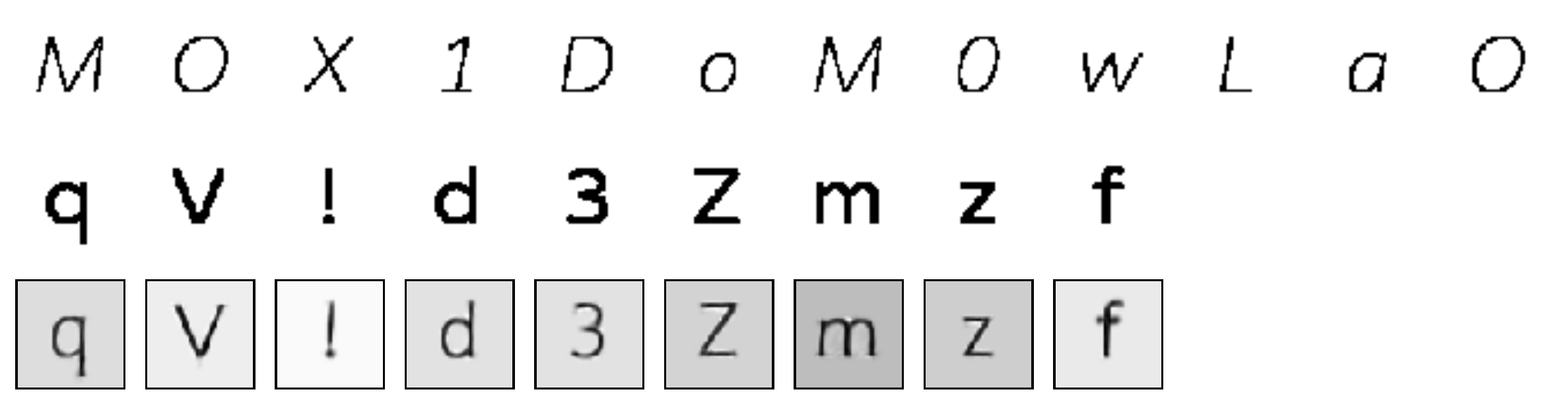}
\caption{First row is domain input, second row is content input, third row is fusion output.}
\end{subfigure}
\caption{Qualitative results of the fusion task on the Google Fonts dataset.}
\label{fig:pairs}
\end{figure}

\begin{figure}
\begin{subfigure}[b]{\textwidth}
\includegraphics[width=\textwidth]{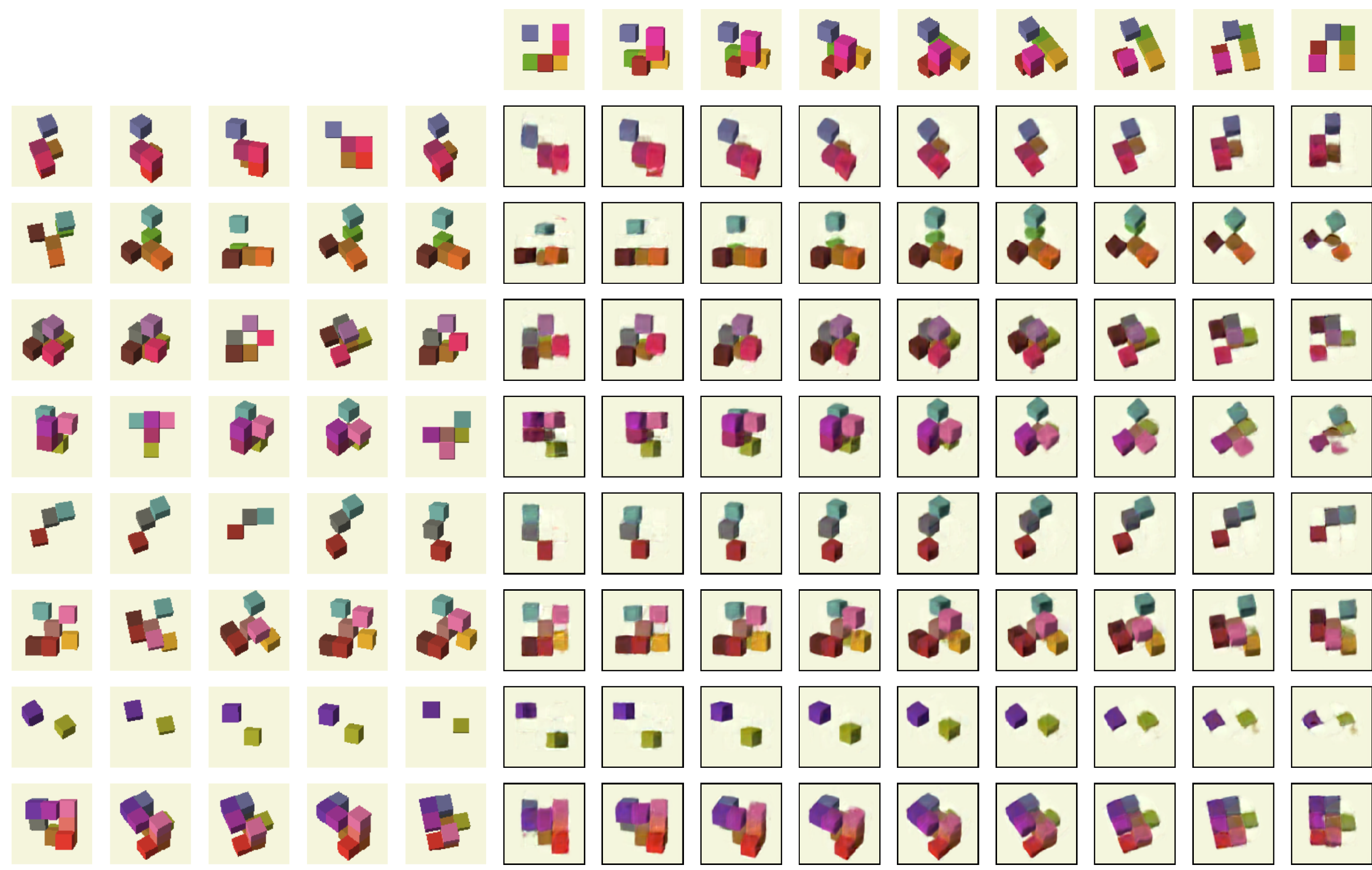}
\end{subfigure}
\\\\
\begin{subfigure}[b]{\textwidth}
\includegraphics[width=\textwidth]{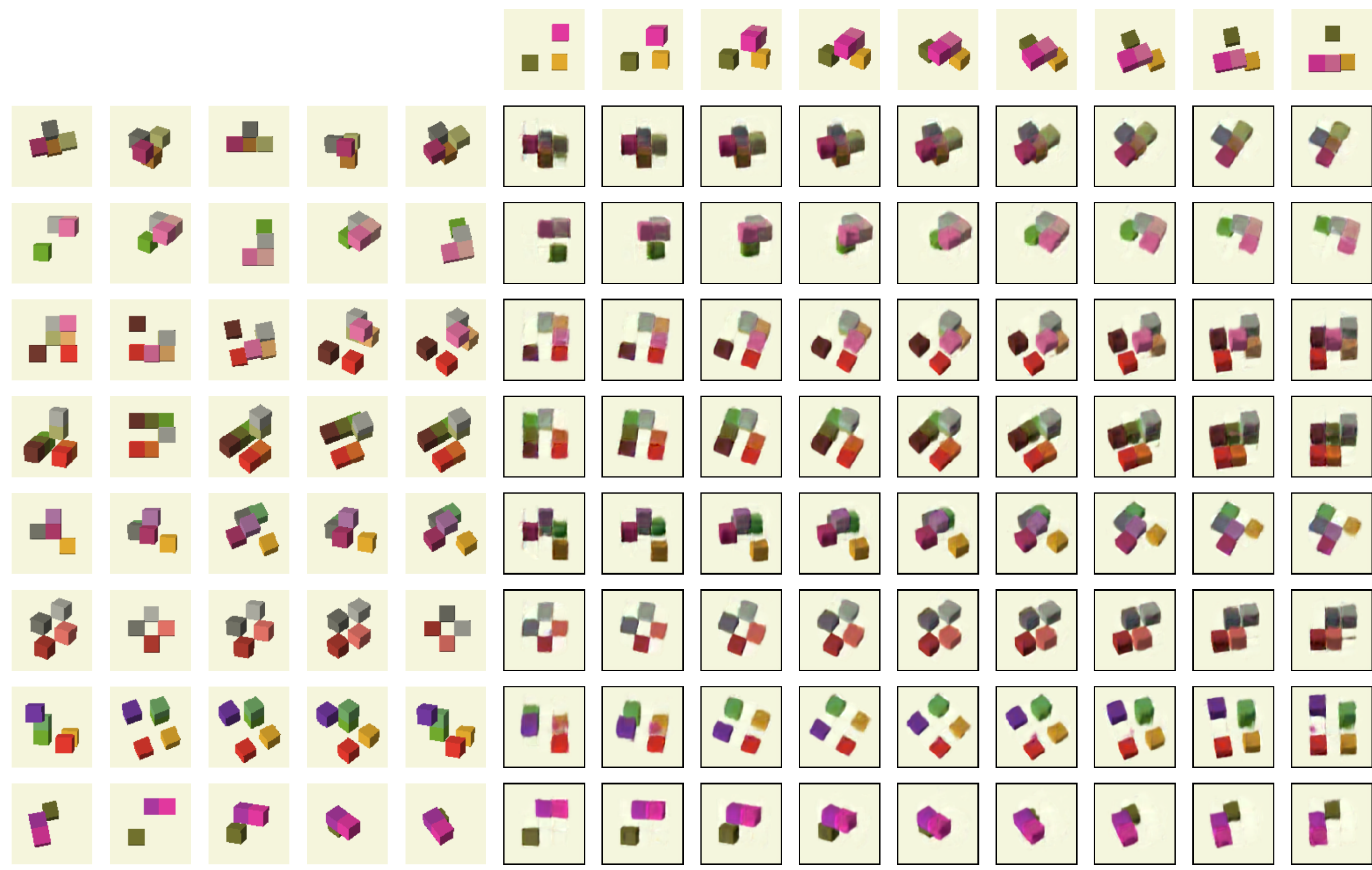}
\end{subfigure}
\caption{Further qualitative results of the fusion task on the Silhouettes dataset. Rows are domains, columns are contents.}
\label{fig:sil}
\end{figure}

Throughout our experiments, the models are trained for 100 epochs and tested only on withheld domains. We provide additional qualitative results on the Google Fonts dataset in Figure \ref{fig:pairs} and on the Silhouettes dataset in Figure \ref{fig:sil}.

\subsection*{Unseen domain and unseen content}

\begin{figure}
\begin{subfigure}[b]{\textwidth}
\includegraphics[width=\textwidth]{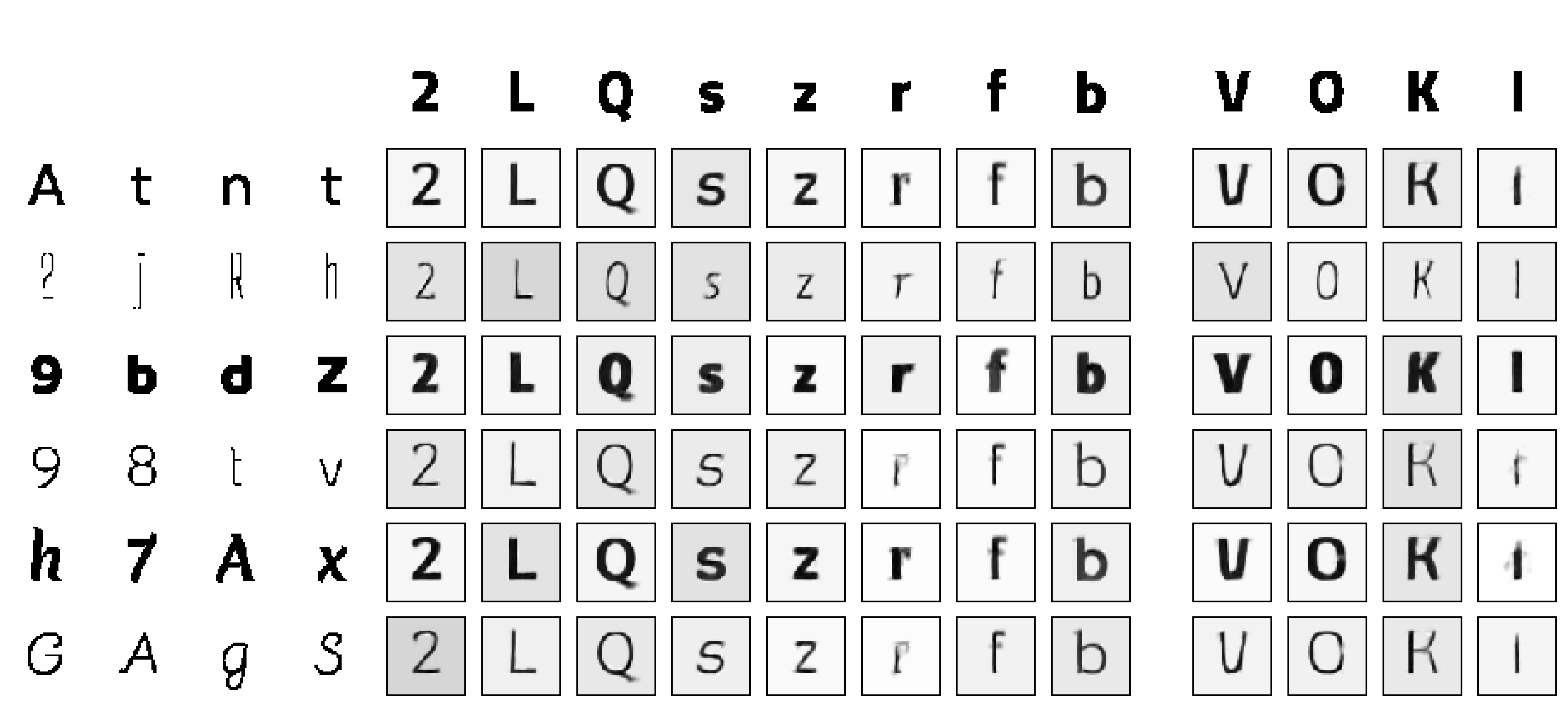}
\end{subfigure}
\\
\begin{subfigure}[b]{\textwidth}
\includegraphics[width=\textwidth]{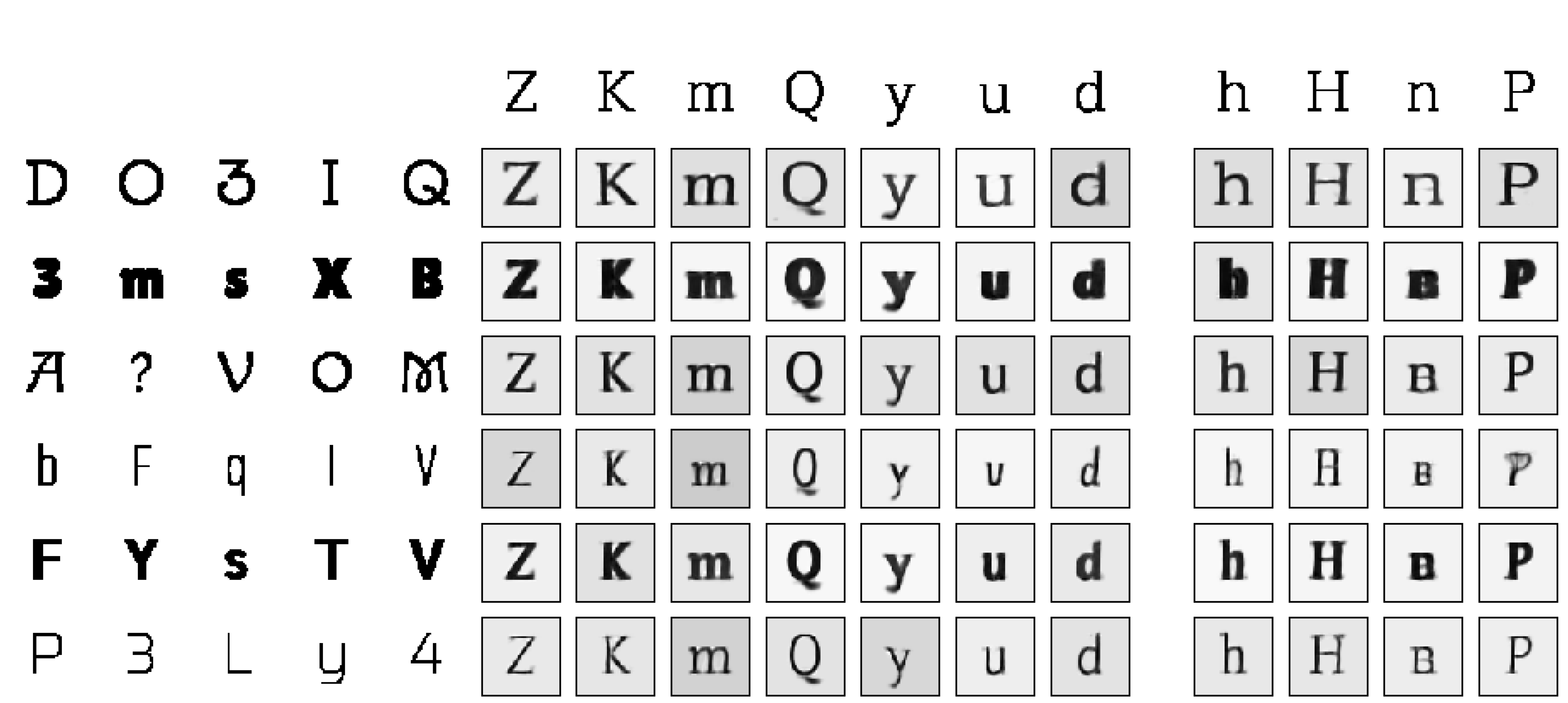}
\end{subfigure}
\\
\begin{subfigure}[b]{\textwidth}
\includegraphics[width=\textwidth]{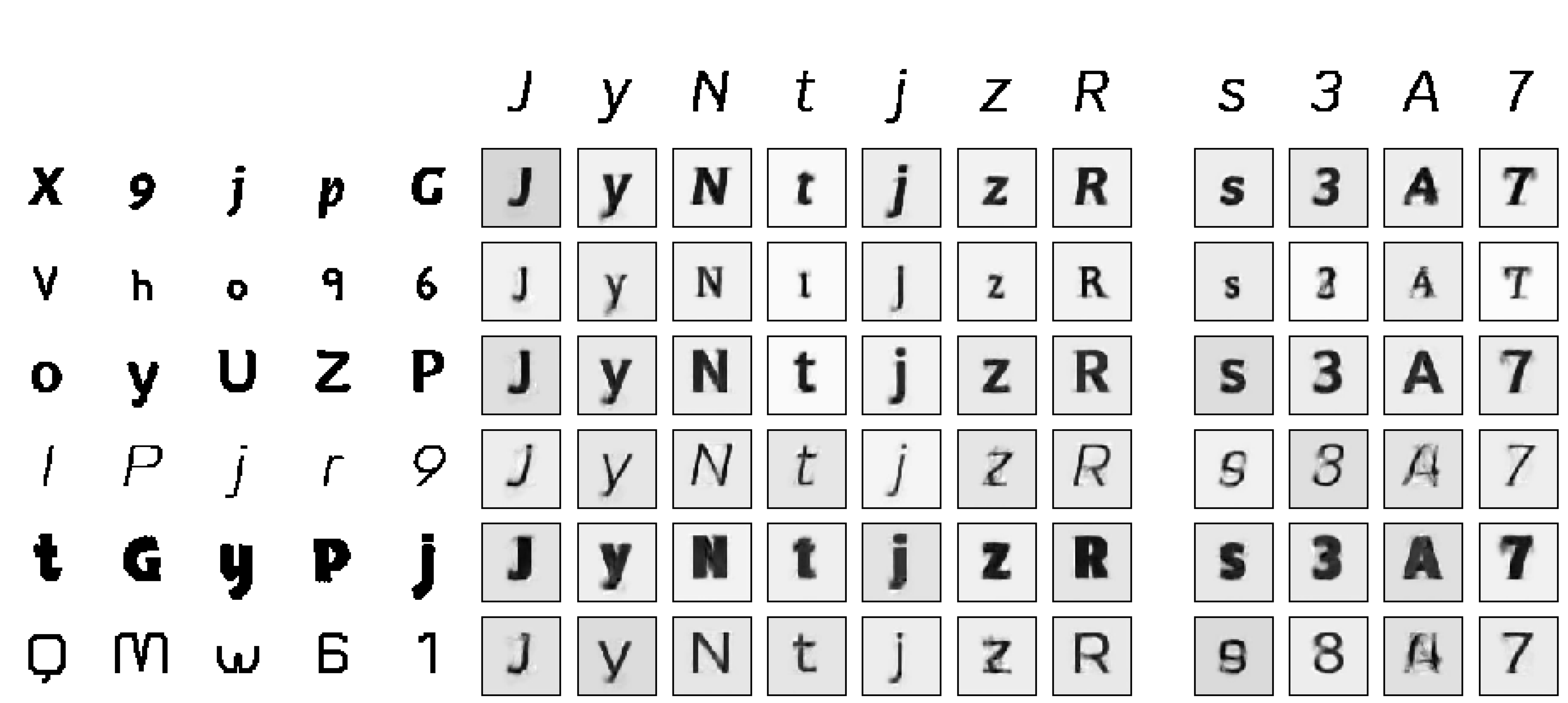}
\end{subfigure}
\caption{Qualitative results of the fusion task with unseen domains and unseen contents. Row is domain, column is content. Leftmost 5 columns are domain inputs. Top rows are content inputs. Center blocks are results with unseen domains - seen contents. The rightmost blocks are results with unseen domains - unseen contents.}
\label{fig:unseen}
\end{figure}

We withhold 4 characters from the Google Font dataset during training, and then attempt the fusion task where both inputs come from unseen domains, and the content input has an unseen content. Results are shown in Figure \ref{fig:unseen}.

\subsection*{Quantitative results}

Table \ref{tab:pred} presents more quantitative results on both the Silhouettes and Google Fonts datasets. We compare both versions of our model (with and without the Domain-Confusion loss) with the VAE \citep{kingma2013auto, rezende2014stochastic}, FactorVAE \citep{kim2018disentangling}, $\beta$-TCVAE \citep{chen2018isolating}, and random guessing. For these models we use the same exact architecture and hyperparameters as with our Domain-Content model. For the FactorVAE we use the same discriminator network as in our model, without the summing and concatenation. We use a $\beta=10$ for the $\beta$-TCVAE and a $\gamma=100$ for the FactorVAE, reflecting the relative weighting that the discriminator loss has in our model. All the latent representations have dimensionality 16.

We obtain the scores by training a 1-layer affine mapping between the latent representation and each data factor for each pre-trained model. The models are trained for 50 epochs, after which the affine mappings are trained for 10 epochs. Each data factor has its own encoding: For the Silhouettes dataset, the domain factor is a 27-way sigmoid classification (one for each cube in the grid), while the content factor is a 2-way regression with values between 0 and 90. The scores are given in average cross-entropy (CE) and mean square error (MSE).

For the Google Fonts dataset, the domain factor is predicted through a verification task whereby a 2-layer MLP with 64 hidden units takes as input two latent representations and has to output 1 if they have the same font or 0 if not. When training, half of the pairs have the same font and half do not. The content factor is a 64-way softmax classification, where each unit represents one character. The scores are given in average cross-entropy.

\begin{table}
  \caption{How well can the latent representation predict the ground-truth factors. Model (I) is the FactorVAE \citep{kim2018disentangling} and model (II) is the $\beta$-TCVAE \citep{chen2018isolating}.}
  \label{tab:pred}
  \centering
  \vspace{0.2cm}
  Silhouettes Dataset
  \\
  \vspace{0.1cm}
  \begin{tabular}{lllllllll}
    \toprule
    & \multicolumn{2}{c}{MO (w/ DC)} & \multicolumn{2}{c}{MO (w/o DC)} \\
    \cmidrule(r){2-5}
    Factor (metric) & domain & content & domain & content & (I) & (II) & VAE & Guess \\
    \midrule
    domain (CE) & \textbf{-0.023} & -0.449 & -0.051 & -0.402 & -0.211 & -0.205 & -0.236 & -0.451 \\
    content (MSE) & 673.4 & \textbf{456.2} & 672.8 & 533.2 & 563.1 & 581.6 & 597.8 & 671.3 \\
    \bottomrule
  \end{tabular}
  \\
  \vspace{0.2cm}
  Google Fonts Dataset
  \\
  \vspace{0.1cm}
  \begin{tabular}{lllllllll}
    \toprule
    & \multicolumn{2}{c}{MO (w/ DC)} & \multicolumn{2}{c}{MO (w/o DC)} \\
    \cmidrule(r){2-5}
    Factor (metric) & domain & content & domain & content & (I) & (II) & VAE & Guess \\
    \midrule
    domain (CE) & \textbf{-0.103} & -0.572 & -0.162 & -0.564 & -0.411 & -0.369 & -0.430 & -0.581 \\
    content (CE) & -0.462 & \textbf{-0.015} & -0.443 & -0.035 & -0.287 & -0.321 & -0.397 & -0.473 \\
    \bottomrule
  \end{tabular}
\end{table}

\end{document}